\documentclass[RNAAS]{aastex62}

\renewcommand{\thefootnote}{\fnsymbol{footnote}}

\newcommand{\gsim}{\lower.7ex\hbox{\ensuremath{\;\stackrel{\textstyle>}{\sim}\;}}}
\newcommand{\lsim}{\lower.7ex\hbox{\ensuremath{\;\stackrel{\textstyle<}{\sim}\;}}}

\begin{document}

\title{A TESS Search for Distant Solar System Objects: Yield Estimates}

\correspondingauthor{Matthew~J.~Payne}
\email{mpayne@cfa.harvard.edu; matthewjohnpayne@gmail.com}
\author[0000-0001-5133-6303]{Matthew~J.~Payne} 
\affiliation{Harvard-Smithsonian Center for Astrophysics, 60 Garden St., MS 51, Cambridge, MA 02138, USA}

\author[0000-0002-1139-4880]{Matthew~J.~Holman}
\affiliation{Harvard-Smithsonian Center for Astrophysics, 60 Garden St., MS 51, Cambridge, MA 02138, USA}

\author[0000-0001-5449-2467]{Andr{\'a}s~P{\'a}l}
\affiliation{Konkoly Observatory, Research Centre for 
Astronomy and Earth Sciences, Konkoly-Thege M. \'ut 15-17, Budapest, H-1121, 
Hungary}
\affiliation{Department of Astronomy, Lor\'and E\"otv\"os University, 
P\'azm\'any P. stny. 1/A, Budapest H-1117, Hungary}


\section{} 

As the NASA Transiting Exoplanet Survey Satellite (TESS) fulfills its primary mission~\citep{2015JATIS...1a4003R}, it is executing an unprecedented survey of almost the entire sky: TESS's approved extended mission will likely extend sky coverage to $\sim94\%$, including $60\%$ of the ecliptic (R. Vanderspek, \emph{priv. comm.}).

In \citet{2019RNAAS...3..160H} we demonstrated that `digital tracking' techniques can be used to efficiently `shift-and-stack' TESS full frame images (FFIs) and showed that combining $\sim\!1,300$ exposures from a TESS sector gives a $50\%$ detection threshold of $I_C\sim 22.0\pm0.5$, raising the possibility that TESS could discover the hypothesized Planet Nine~\citep{Trujillo.2014,Brown.2016,Fortney.2016,2019arXiv190210103B}.
We note that the threshold realized in practice may skew brighter (e.g. $I_C\gsim 21.5$), due to the combined effects of stellar contamination and un-modelled stray light (L. Bouma, \emph{priv. comm.}).

The practical constraint on this technique is the number of mathematical operations to be performed when trying all plausible orbits:
\begin{eqnarray}
N_{\rm op}  & = &  N_{\rm sec} N_{\dot\alpha}  N_{\dot\beta} N_{\gamma} N_{\rm pix} N_{\rm exp} \nonumber \\ 
& \sim & 
2\times10^{16}
\left(\frac{N_{\rm sec}}{26\,\mathrm{sectors}}\right)
\left(\frac{T}{27\,\mathrm{day}}\right)^3
\left(\frac{P}{21''}\right)^{-3}
\left(\frac{d}{25\,{\rm au}}\right)^{-4}
\left(\frac{N_{\rm exp}}{1,300}\right)
\left(\frac{N_{\rm pix}}{16\,\mathrm{Mpix}}\right),
\label{e:Nop}
\end{eqnarray}
where 
$N_{\rm sec}$ is the number of sectors, 
$N_{\dot\alpha}$ and $N_{\dot\beta}$ the number of angular velocity bins to be searched, 
$N_{\gamma}$ the number of distance bins, 
$T$ the observational span, 
$P$ the sky-plane resolution, 
$d$ the distance, 
$N_{\rm exp}$ the number of exposures,  
and 
$N_{\rm pix}\propto P^{-2}$ the number of pixels (see \citet{2019RNAAS...3..160H} for more details\footnote{Eqn 1 properly accounts for all 26 sectors, correcting an error in \citet{2019RNAAS...3..160H}.}).
The successful \citet{Bernstein04} search of HST data required $N_{\rm op}\sim10^{16}$.

In this note, we demonstrate that this technique has the  potential to discover hundreds of Kuiper Belt Objects (KBOs) and Centaurs in TESS FFI data.

\section*{Very Distant Objects}

Distant objects (e.g. Planet-9) must move sufficiently within the span of observations to be distinguishable from stationary, background sources.  
Assuming a minimum detectable displacement of $n_p\sim5$ pixels\footnote{TESS pixels are $\sim21''$ across}, TESS can detect moving objects at distances of $d \lesssim 900\left(\frac{5\,\mathrm{pix}}{n_{p}}\right)\,au$. 
At this distance, the number of operations required to exhaustively search a sector becomes trivially small ( $N_{op}\sim10^{10}$) due to the small number of physically plausible pixel-shifts required.

\section*{Kuiper Belt Objects (KBOs)}
\begin{figure*}
    \begin{minipage}[b]{\textwidth}
    \centering
    \includegraphics[angle=0,width=0.99\columnwidth]{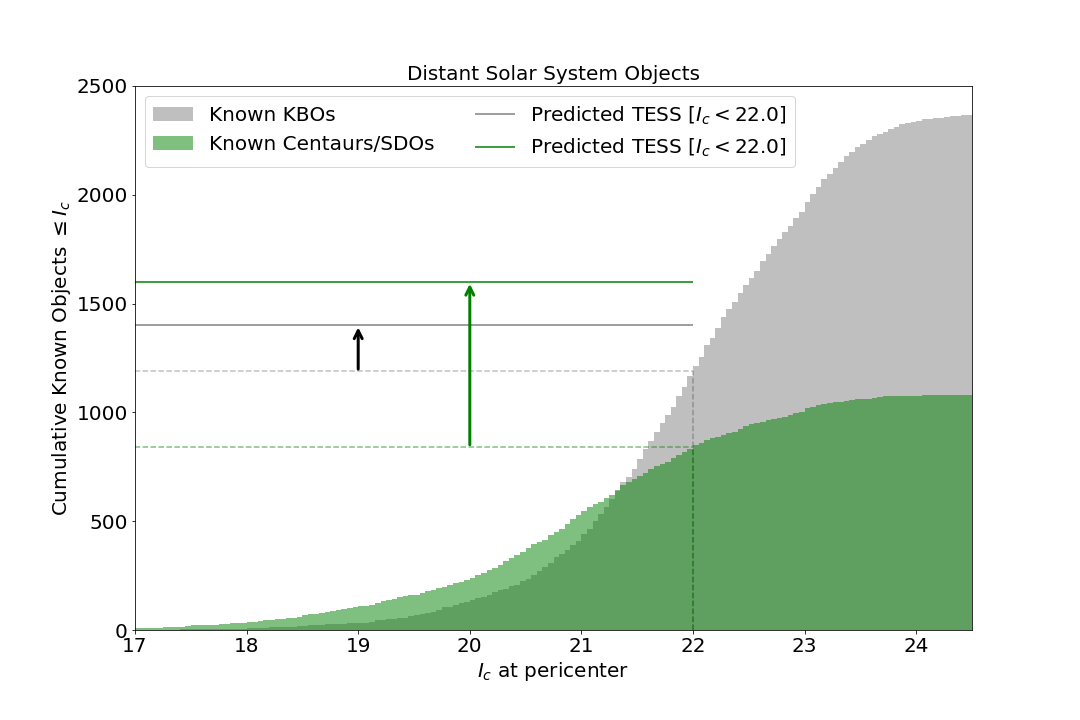}
    \caption{
    Cumulative histograms of known KBOs (gray) and Centaurs/SDOs ({\textcolor{green}{green}}) are plotted as functions of $I_c$ (at pericenter), along with dashed lines indicating the current known numbers with $I_C<22.0$.
    Our predicted TESS yields for $I_C<22.0$ are plotted as solid horizontal lines, indicating significant numbers of expected discoveries. 
} 
    \label{f:known}
    \end{minipage}
\end{figure*}

Eqn. \ref{e:Nop} indicates that a KBO search will require $N_{op}\sim10^{16}$ for the entire nominal TESS mission, similar to the \citet{Bernstein04} search, but we emphasize that CPUs are now at least $10^3$ cheaper per GFlop\footnote{https://en.wikipedia.org/wiki/FLOPS} than in 2004. 
The surface density of KBOs near the ecliptic brighter than $I_C = 22.0$ is $\sim 0.2\,\mathrm{deg}^{-2}$~\citep{Gladman.2001,2014ApJ...782..100F}, 
where we have assumed average colors of $V-R\sim0.6$ and $R-I_C\sim0.6$ for these objects~\citep{2015A&A...577A..35P}.  
Most KBOs orbit within $12\arcdeg$ of the ecliptic~\citep{Brown.2001}, and we estimate that by the end of its extended mission, TESS will have observed $\sim 80\%$ of this region, for an effective area of $7,000\,\mathrm{deg}^2$
yielding $\sim 1,400$ KBOs brighter than $I_C = 22.0$ within 12 degrees of the ecliptic. 
In Figure \ref{f:known} we compare the known population of KBOs with the predicted yield, demonstrating that TESS has the potential to discover $\sim200$ \emph{new} KBOs with $I_c\geq22.0$.

\section*{Centaurs}

A TESS search for Scattered Disk Objects (SDOs) and Centaurs (with $d\gsim10\, au$) is also computationally feasible ($N_{op} \sim 4\times10^{17}$).  
For Centaurs and SDOs, the combined surface density is $\sim 0.1\,\mathrm{deg}^{-2}$~\citep{2001AJ....121..562L}, but they are more broadly distributed in ecliptic latitude than are KBOs.  
Their latitude distribution is poorly known due to the comparatively small area of previous surveys. 
Conservatively, we assume a discovery area twice that for KBOs, yielding $\sim1,600$ objects.  
In Figure \ref{f:known} we compare the known population of Centaurs and SDOs with our predicted yield, demonstrating that many remain to be discovered. 
TESS has the potential to discover $\sim750$ \emph{new} Centaurs/SDOs with $I_c\geq22.0$.

\section*{Other}

Searching for objects with $d\ll10\, au$ becomes increasingly challenging because $N_{op}\propto\,d^{-4}$.
For the asteroid belt, $N_{op}\sim10^{18}$, which is likely infeasible. 
However, one could restrict a search to (e.g.) prograde orbits within $\pm5^{\arcdeg}$ of the ecliptic, reducing the number of angular velocity bins to be searched, $N_{\dot\alpha}\times N_{\dot\beta}$, (and hence $N_{op}$) by a factor of 30. 

Alternatively, bright objects such as the interstellar comet \emph{2I/Borisov}, currently around 17$^{th}$ magnitude, need little stacking, facilitating a well-characterized all-sky survey to constrain their occurrence-rate.

\section*{Acknowledgments}
We acknowledge 
Charles Alcock,  
Michele Bannister,  
Tom Barclay,
Geert Barentsen, 
Luke Bouma,
Jason Eastman, 
Adina Feinstein,
Wes Fraser, 
Pedro Lacerda, 
Ben Montet,
George Ricker, 
Scott Sheppard, 
David Trilling,
Chad Trujillo, 
Roland Vanderspek, 
Joshua Winn,
Deborah Woods, 
and grants from NASA (NNX16AD69G, NNX17AG87G, UM-Subaward 46039-Z6110001), Hungary (NKFIH-K125015), and 
the Smithsonian.

\bibliographystyle{aas-compact}


\end{document}